\documentclass{appolb}
\usepackage{graphicx}
\usepackage{dirtytalk}
\usepackage{lineno}

\begin{document}
\title{Recent results from STAR for parton distribution functions at low and high $x$ in proton-proton collisions%
\thanks{Presented at ``Diffraction and Low-$x$ 2024", Trabia (Palermo, Italy), September 8-14, 2024.}%
}
\author{Zilong Chang for the STAR collaboration
\address{Center for Exploration of Energy and Matter, Indiana University \\ Bloomington, Indiana, USA 47408}
%\\[3mm]
%{Third Author % of different affiliation
%\address{affiliation}
%}
%\\[3mm]
%the Name(s) of other Author(s)
%\address{affiliation}
}
\maketitle
\begin{abstract}
%Jets, clusters of collimated particles produced in high energy proton-proton ($pp$) collisions, serve as a useful channel for studying the internal structure of the proton.
%the center-of-mass energies of
According to perturbative quantum chromodynamic calculations, in $pp$ collisions at $\sqrt{s} = $ 200 and 510 GeV studied at RHIC, jet production in mid-pseudorapidity, $|\eta| <$ 1, is dominated by quark-gluon and gluon-gluon scattering processes. Therefore  jets at RHIC are direct probes of the gluon parton distribution functions (PDFs) for momentum fractions 0.01 $<x<$ 0.5. Moreover, $W$ boson cross-section ratio, $\sigma(W^+)/\sigma(W^-)$, in $pp$ collisions at $\sqrt{s} = 510$ GeV, is an effective tool to explore anti-quark PDFs, $\bar{d}/\bar{u}$. Last but not least, di-$\pi^{0}$ correlation in forward pseudorapidity, $2.6 < \eta <4.0$, is an important indication of the non-linear gluon dynamics at low $x$ where the gluon density is high in protons and nuclei. In this proceeding, we present recent STAR results of mid-pseudorapidity inclusive jet cross-sections at $\sqrt{s} =$~200 and 510 GeV in $pp$ collisions, $W$ boson cross-section ratio at $\sqrt{s} = 510$~GeV in $pp$ collisions, and forward di-$\pi^0$ correlations in $pp$, $p\textrm{Al}$ and $p\textrm{Au}$ collisions at $\sqrt{s_{\textrm{\tiny NN}}} = 200$~GeV.
% The theoretical implications of these results will also be discussed.
\end{abstract}
  
\section{Introduction}
%The internal structure of the proton has been studied extensively through a series of high energy scattering experiments.
As the wavelength of a probe decreases below a Fermi ($10^{-15}$~m) in high energy scattering experiments, the image of the proton evolves from a naive picture of three quarks, $uud$, to a complex system of quarks, anti-quarks and gluons known as the parton model. Parton distribution functions (PDFs), interpreted as probabilities of finding a parton carrying a fractional momentum of the proton, $x$, at the probe scale, $Q^2$, are crucial to describe the proton structure. By colliding proton-proton beams at $\sqrt{s} =$~200 and 510 GeV, the STAR experiment~\cite{star} has provided a unique venue to unravel the proton's PDFs since the 2000s. In this proceeding, three important measurements at STAR, inclusive jet cross sections, $W^{\pm}$ boson cross-section ratio, and di-$\pi^0$ correlation in the forward region, are discussed.%, each of which yields unique and critical information.
\section{Inclusive jet cross sections}
Jets are defined as clusters of collimated final-state particles. For each production channel, the cross-sections can be factorized into the PDFs and partonic scattering cross-sections. The partonic scattering cross-sections are calculated perturbatively at a large $Q^2 \gg \Lambda_{\textrm{QCD}}^2$. Due to their non-perturbative nature, the PDFs are parametrized as a function of $x$ at an initial $Q_0^2$ scale and then evolved to the proper $Q^2$. Given their universality, the parameters are determined by fits to global experimental data.

Recent global analyses at next-to-next-to-leading order (NNLO) with TeV-scale $p\bar{p}$ and $pp$ data from the Tevatron and LHC, respectively, yielded gluon PDFs at $Q=100$~GeV with small uncertainties for $x$ from $10^{-4} $ to $0.1$, but with large uncertainties especially for $x > 0.2$~\cite{nnpdf}. At RHIC energies, $\sqrt{s} =$~200 and 510 GeV, the jet production is dominated by quark-gluon and gluon-gluon scatterings~\cite{jetpaper}. Therefore jet cross sections are ideal to constrain the gluon PDFs at $x > 0.1$.
% high momentum $p_T$ jet events are triggered by the fast-response electro-magnetic calorimeters 

At STAR, jets are identified from tracks reconstructed by the Time Projection Chamber and energy deposits in the towers ($0.05 \times 0.05$ in $\eta$-$\phi$) of the electro-magnetic calorimeters (EMC)~\cite{jetpaper}.  We chose to use the anti-$k_T$ algorithm with jet parameter $R=$~0.6 and 0.5 at $\sqrt{s} =$~200 and 510 GeV, respectively. The smaller $R$ at the higher $\sqrt{s}$ is to reduce the contributions from pile-up events and soft backgrounds.

The inclusive jet cross sections at $\sqrt{s} =$~200 and 510 GeV were analyzed from data collected in 2012. Compared to the previous analysis from the 2006 data, a number of technical advancements were made to improve the systematic uncertainties. One is an off-axis cone method to correct for the underlying event (UE) contribution to the jet $p_T$~\cite{jetpaper}.  Another is an improved event generator tune based on the Pythia6 Perugia 2012 tune~\cite{jetpaper}. The simulation is essential in order to unfold the jet cross sections from the measured detector-level jet $p_T$ and $\eta$ bins to the true particle-level jet $p_T$ and $\eta$ bins. In addition, it is necessary to estimate hadronization corrections needed to compare with theoretical calculations at the parton-level.

\begin{figure}[htb]
\centering
 \begin{minipage}{0.48\textwidth}
\centerline{%
\includegraphics[width=\columnwidth]{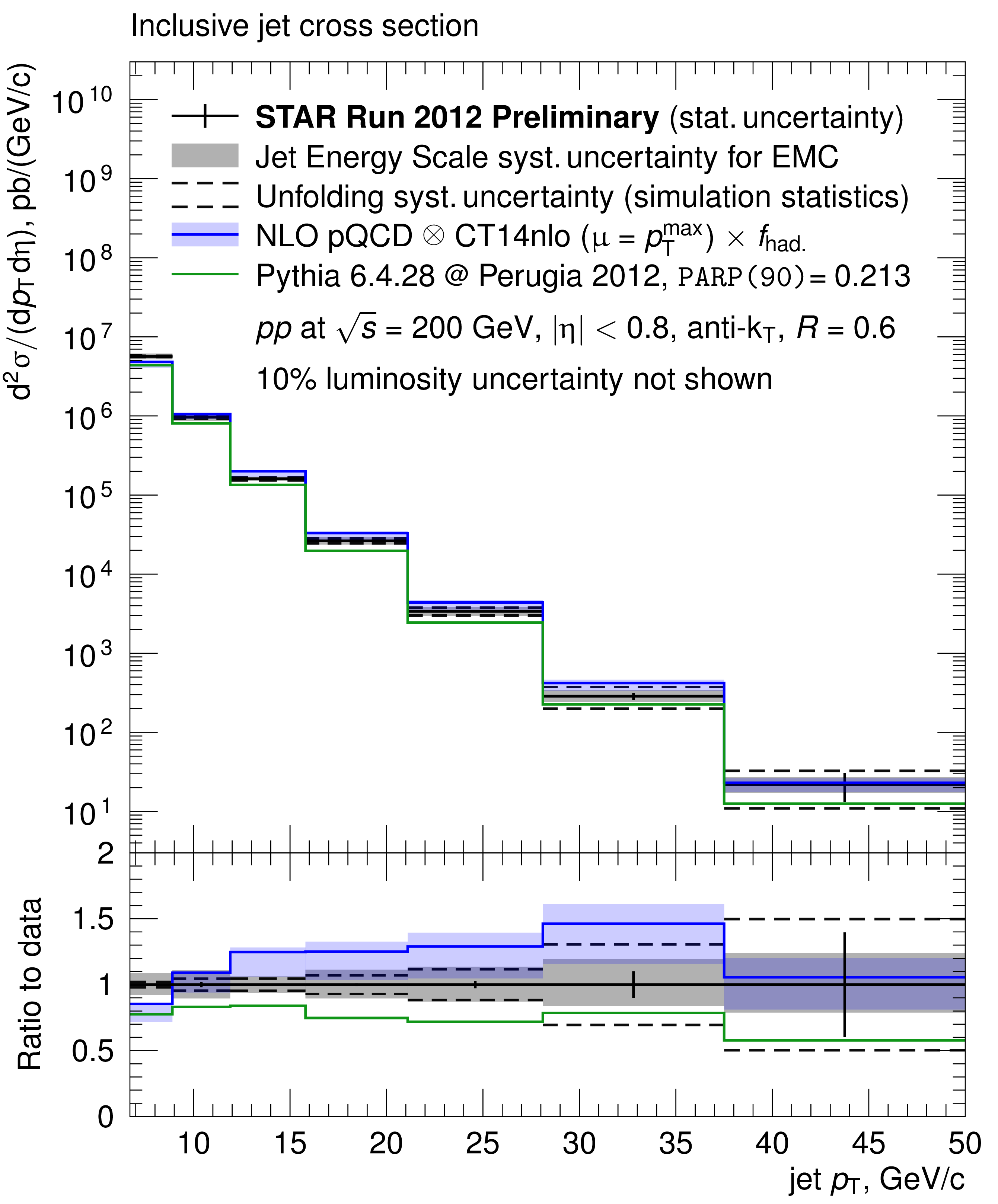}
}
\end{minipage}
\begin{minipage}{0.48\textwidth}
\centerline{%
\includegraphics[width=\columnwidth]{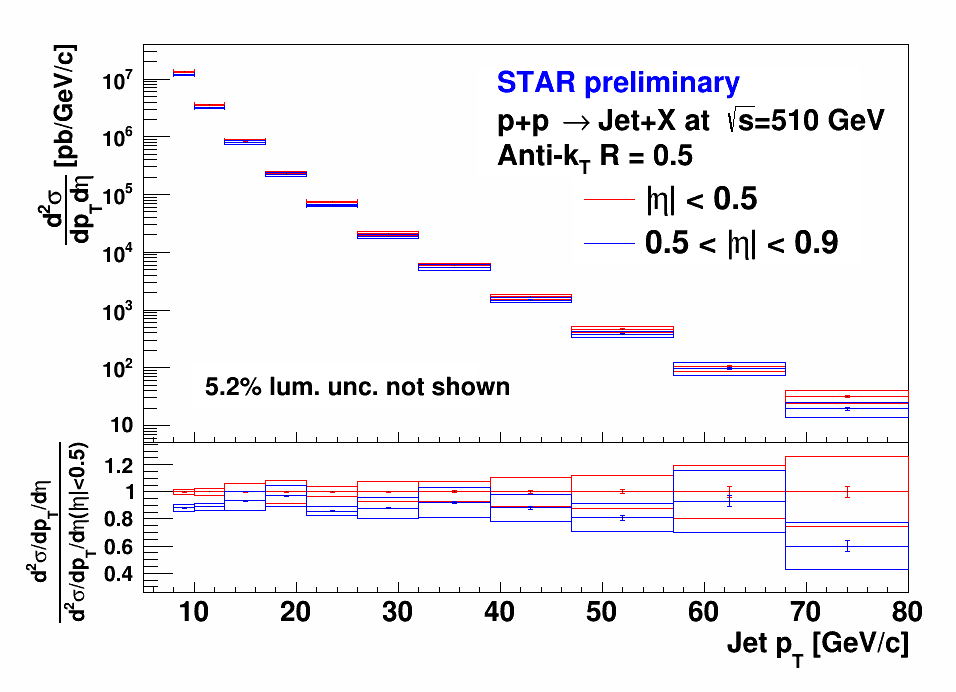}
}
\end{minipage}
\caption{STAR preliminary inclusive jet cross section $\frac{d^2\sigma}{dp_Td\eta}$ vs. particle jet $p_T$ after UE corrections from $pp$ collisions at $\sqrt{s} =$~200 (left) and 510 (right) GeV~\cite{jetprelim}.}
\label{Fig:jetcrs}
\end{figure}

The preliminary results of the inclusive jet cross sections, $\frac{d^2\sigma}{dp_Td\eta}$, as a function of particle jet $p_T$ after UE corrections are presented in Fig. \ref{Fig:jetcrs}~\cite{jetprelim}. The 200 GeV results cover $|\eta|<0.8$, and the 510 GeV results are split into $|\eta| < 0.5$ and $0.5 < |\eta| < 0.9$. The luminosities are determined from Van der Meer scans~\cite{vernier} with uncertainties of 10\% and 5.2\% at 200 and 510 GeV, respectively. The dominant systematic uncertainties come from the jet energy scale for both results. The results sit about 20\% below the perturbative calculations with the CT14 next-to-leading order (NLO) PDF after hadronization corrections. Although the results match the shape of the Pythia6 predictions quite well, an overall scale difference of 20\% is observed.

%\subsection{Subsection}
\section{$W^{\pm}$ cross-section ratio}
The New Muon Collaboration discovered that there were more $d$-flavor sea quarks, $\bar{d}$, than $u$-flavor sea quarks, $\bar{u}$, in the proton~\cite{nmc}. This spurred interest in studying the sea quark distributions from $pp$ collisions through the Drell-Yan (DY) process. The recent SeaQuest experiment, in which proton beams impinged on fixed hydrogen and deuterium targets, demonstrated that $\bar{d}/\bar{u} > 1$ as a function of $x$~\cite{seaquest}. However, the results showed a different trend at high $x$ than the previous NuSea results. Several theoretical predictions have been proposed to explain both trends at high $x$ from these two results.

In $pp$ collisions at $\sqrt{s} =$~510 GeV, $W^{\pm}$ can be produced by the $s$-channel from quark and anti-quark interactions~\cite{starwr}. $W^{\pm}$ cross-section ratio, $R_W = \sigma(W^+)/\sigma(W^-)$, is proportional to $\bar{d}/\bar{u}$ at leading order. Unlike the DY process, $R_W$ is sensitive to the sea quark PDFs at $Q^2 = M_W^2$, where $M_W^2$ is the invariant mass of the $W$ boson. The sampled $x$ range is $0.06 < x < 0.4$ when $-1.0 < \eta_{W} < 1.5$.

At STAR, $W^{\pm}$s are reconstructed from the high-energy decay $e^{\pm}$~\cite{starwr}. The $e^{\pm}$ candidates are captured by clusters of EMC towers, each of which spans $0.1 \times 0.1$ in $\eta$-$\phi$. A large $p_T$ imbalance is required in the event to account for the missing final-state neutrino. Corrections are made for backgrounds from electroweak residuals and the QCD dijet contributions.

\begin{figure}[htb]
\centerline{%
\includegraphics[width=0.9\columnwidth]{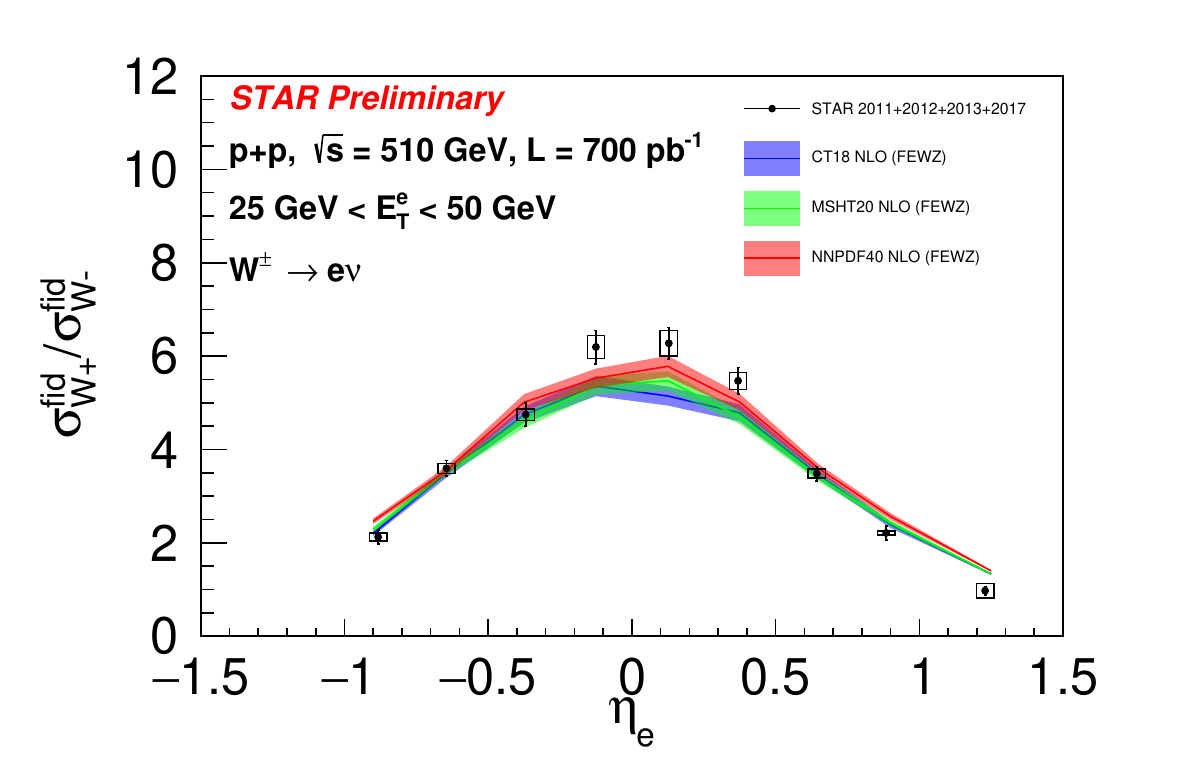}
}
\caption{STAR preliminary $W^{\pm}$ cross-section ratio, $R_W$, vs. reconstructed $\eta_e$ from $pp$ collisions at $\sqrt{s} =$~510 GeV~\cite{wrprelim}.}
\label{Fig:WRatio}
\end{figure}

The preliminary results of $R_W$ as a function of the reconstructed $\eta$ of $e^{\pm}$, are shown in Fig. \ref{Fig:WRatio}~\cite{wrprelim}. The plot includes data collected in 2011, 2012, 2013, and 2017 with total integrated luminosity of 700 $\textrm{pb}^{-1}$. The results agree well with the recent predictions from the NLO PDF sets, such as CT18~\cite{ct18}, MSHT20~\cite{msht20}, and NNPDF4.0~\cite{nnpdf}. Note that the NNPDF4.0 PDF has incorporated the recent SeaQuest data.% in their analysis.
\section{Di-$\pi^0$ correlations in the forward region}
High energy physics experiments have shown that gluons dominate at low $x$ inside the proton, and the number of gluons grows as $x$ decreases~\cite{nnpdf,ct18,msht20}. An interesting theoretical expectation is that the gluon density must begin to saturate at a certain $x$ where the rate of gluon splittings balances out gluon recombinations. The probe scale where the saturation happens, $Q_S^2$, varies inversely as a function of $x$. In heavy nuclei, $Q_S^2$ is proportional to $A^{1/3}$, where $A$ is the atomic number of a nucleus.

The color glass condensate (CGC) framework predicts a suppression and an azimuthal broadening of the back-to-back di-hadrons in $p\textrm{A}$ collisions compared to $pp$ when gluon saturation appears~\cite{stardipi0}. The di-$\pi^0$ azimuthal correlation in $2.6 < \eta < 4.0$ allows one to study gluons with $x$ as low as $10^{-4}$ in $pp$, $p\textrm{Al}$ and $p\textrm{Au}$ collisions at $\sqrt{s_{\textrm{\tiny NN}}} = $ 200 GeV.

\begin{figure}[htb]
 \begin{minipage}{0.48\textwidth}
\centerline{%
\includegraphics[width=\columnwidth]{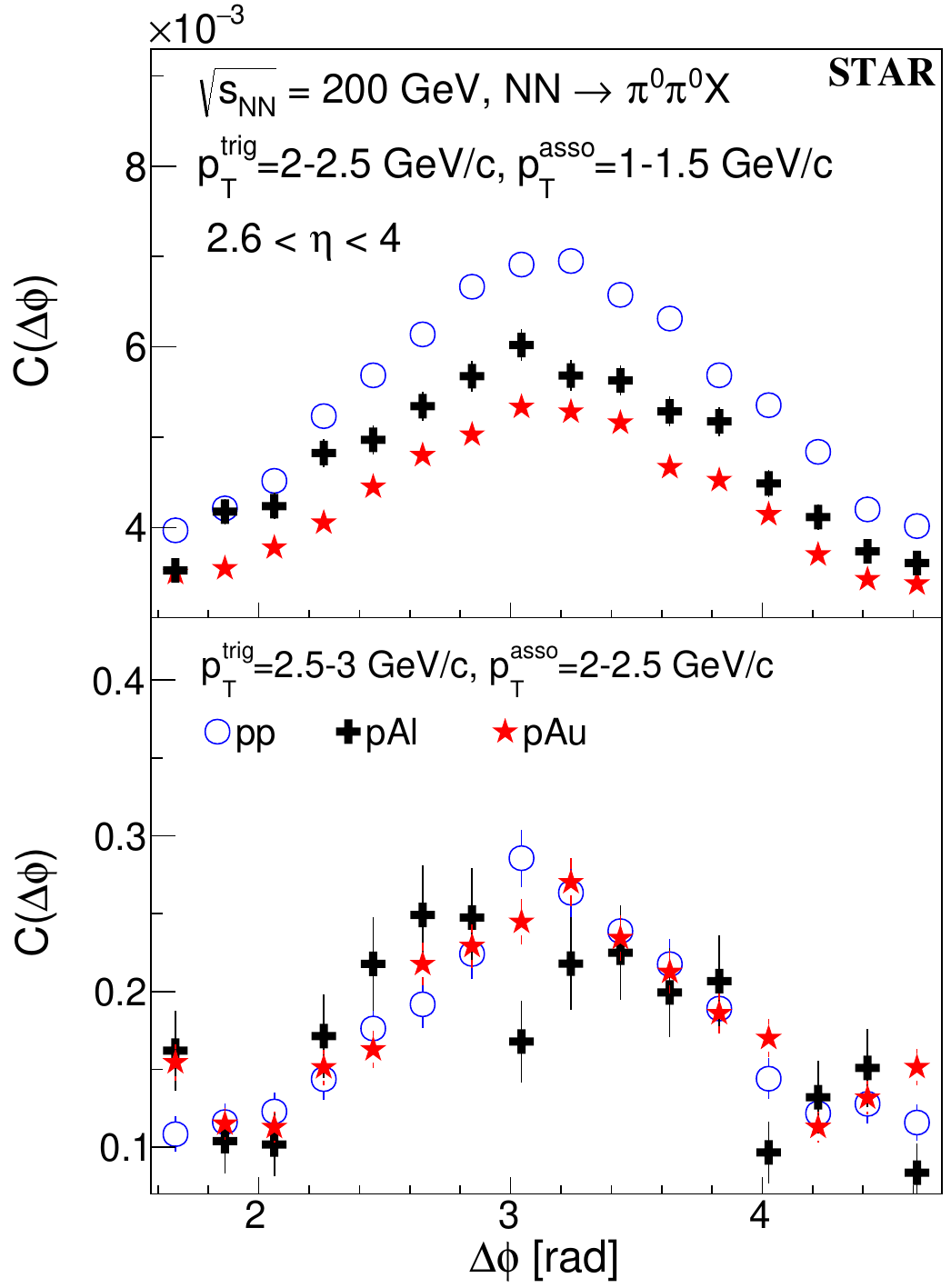}
}
\end{minipage}
 \begin{minipage}{0.48\textwidth}
\centerline{%
\includegraphics[width=\columnwidth]{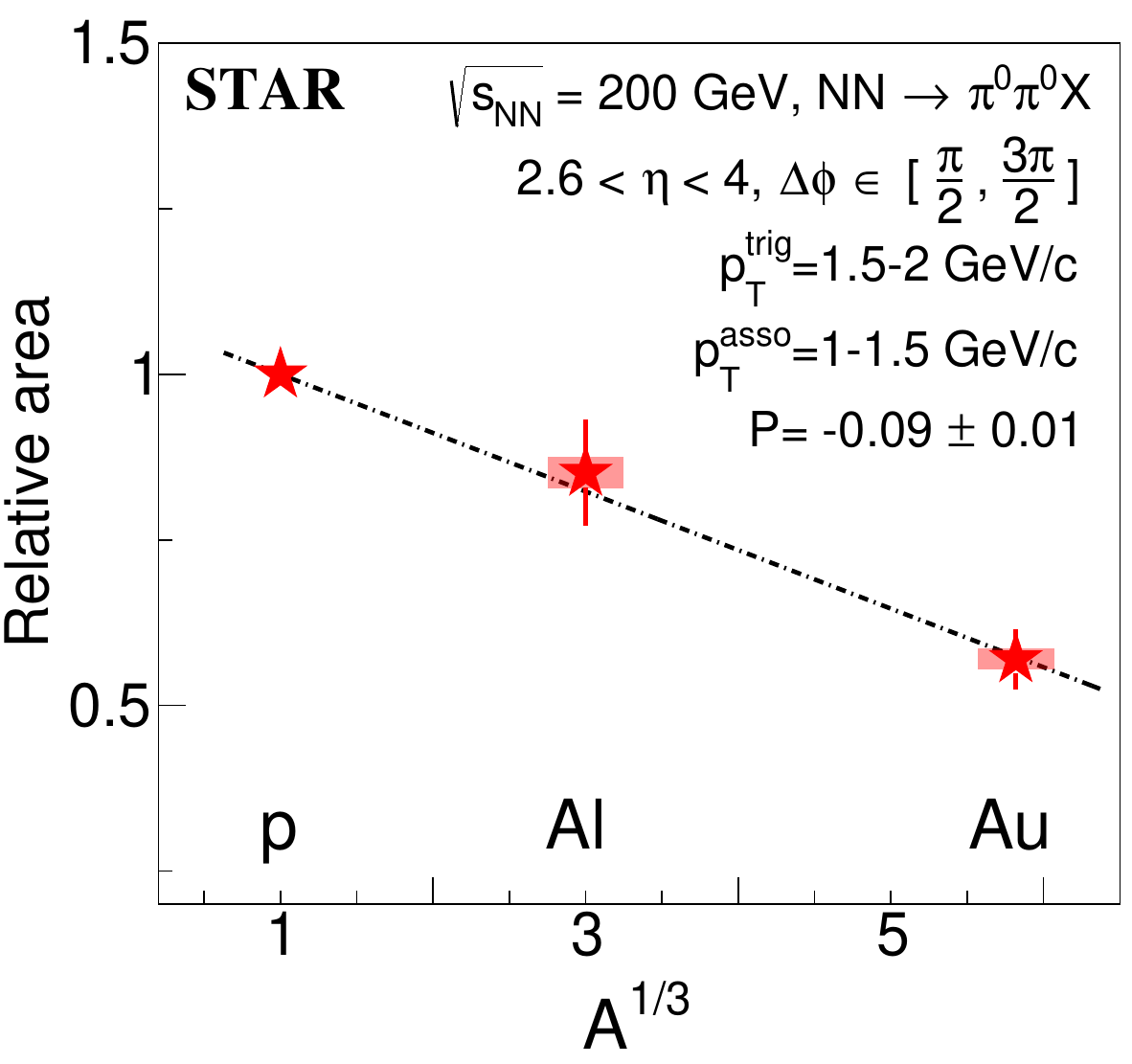}
}
\end{minipage}
\caption{STAR di-$\pi^0$ correlation function $C(\Delta \phi)$ in selected $p_T$ ranges of trigger and associate $\pi^0$s from $pp$, $p\textrm{Au}$ and $p\textrm{Al}$ collisions at $\sqrt{s_{\textrm{\tiny NN}}} =$~200 GeV on the left panel, and the corresponding relative areas on the right panel~\cite{stardipi0}.}
\label{Fig:dipionArea}
\end{figure}

At STAR, $\pi^0$s can be reconstructed from decay photons detected in the Forward Meson Spectrometer (FMS). A correlation function in terms of di-$\pi^0$ opening angle, $\Delta \phi$, can be defined by $C(\Delta \phi) = \frac{N_{\textrm{pair}}(\Delta \phi)}{N_{\textrm{trig}} \times \Delta \phi_{\textrm{bin}}}$, where $N_{\textrm{pair}}(\Delta \phi)$ is the number of di-$\pi^0$ pairs in a $\Delta \phi$ bin, $\Delta \phi_{\textrm{bin}}$ is the $\Delta \phi$ bin width,  and $N_{\textrm{trig}}$ is the number of trigger $\pi^0$s. The trigger $\pi^0$ is the higher $p_T$ of the pair, and the associate $\pi^0$ is the lower $p_T$ one. $C(\Delta \phi)$ extracted from the 2015 STAR $pp$, $p\textrm{Al}$, and $p\textrm{Au}$ collisions is shown in Fig. \ref{Fig:dipionArea}~\cite{stardipi0}.

As plotted in Fig. \ref{Fig:dipionArea}, the relative area under $C(\Delta \phi)$ from $\Delta \phi = \pi/2$ to $3\pi/2$ comparing to the $pp$ collisions, follows linearly as a function of $A^{1/3}$. The kinematic choices with the trigger $\pi^0$, $p_T^{\textrm{trig}} =$~1.5-2 $\textrm{GeV}/c$, and the associate $\pi^0$, $p_T^{\textrm{asso}} =$~1-1.5 $\textrm{GeV}/c$ are where gluon density is large and expected to saturate. Surprisingly no angular broadening is observed.
\section{Forward upgrade}
The STAR forward upgrade~\cite{rhicqcd2023} adds new capabilities in place of the FMS over a similar $\eta$ range, $2.6 < \eta < 4.0$. A new hadronic calorimeter is located behind an upgraded EMC. Together they constitute the forward calorimeter system. In addition, a forward tracking system consists of a silicon tracker and a small thin gap gas chamber. The forward upgrade has the capability of separating the charge signs of charged hadrons. The detectors were commissioned in 2022 and have been successfully taking data since then. The forward upgrade enables STAR to study asymmetric partonic collisions, where $x_1 \gg x_2$, therefore one can explore both high-$x$ and low-$x$ regimes. In particular it allows one to study valence quark distributions as high as $x > 0.5$ where no current experiment has reached.
%
%\begin{figure}[htb]
%\centerline{%
%\includegraphics[width=10cm]{xQ2fwd.png}
%}
%\caption{$x$-$Q^2$ coverage with the STAR forward upgrade detector}
%\label{Fig:dipionArea}
%\end{figure}

%uncomment the following lines to place a figure
\section{Conclusion}
STAR has a vigorous physics program centered around exploring the internal structure of the proton. Inclusive jet production is sensitive to the large $x$ gluon PDFs in the proton that are loosely constrained by data from TeV-scale colliders. $W^{\pm}$ cross-section ratio is complementary to the DY process to constrain $\bar{d}/\bar{u}$. Di-$\pi^0$ correlation at $2.6 < \eta < 4.0$ in $pp$ and $p\textrm{A}$ collisions enables the study of the non-linear dynamics of gluons at low $x$. With upcoming exciting physics results from the forward upgrade, STAR will lay the essential groundwork for the future Electron Ion Collider.


\begin{thebibliography}{99}
\bibitem{star}
K.~H.~Ackermann {\it et al.},
%\emph{STAR detector overview},
\emph{Nucl. Instr. and Meth. A} {\bf 499} 624 (2003).
\bibitem{nnpdf}
R.~D.~Ball {\it et al.}, (NNPDF)
\emph{EPJC} {\bf 82} (2022), 428
\bibitem{jetpaper}
J.~Adam {\it et al.}, (STAR)
%\emph{Longitudinal double-spin asymmetry for inclusive jet and dijet production in $pp$ collisions at $\sqrt{s}=$ 510 GeV}
\emph{Phys. Rev. D} {\bf 100} 052005. (2019).
%\bibitem{antikt}
%M.~Cacciari, G.~Salam, and G.~Soyez,
%\emph{The anti-$k_t$ jet clustering algorithm},
%\emph{JHEP} {\bf 0804} 063 (2008).
\bibitem{jetprelim}
Z.~Chang for the STAR Collaboration
\emph{PoS}(PANIC2021)377.
\bibitem{vernier}
S.~Van~Der~Meer,
%\emph{Calibration of the effective beam height in the ISR},
\emph{ISR-PO/68-31} (1968).
%\bibitem{perugia}
%P. Skands,
%\emph{Tuning Monte Carlo generators: The Perugia tunes},
%\emph{Phys. Rev. D} {\bf 82} 074018 (2010) [{\tt arXiv:1005.3457v5[hep-ph]}].
%\bibitem{pythia}
%T. Sjostrand, S. Mrenna, and P. Skands,
%\emph{PYTHIA6.4 physics and manual},
%\emph{JHEP} {\bf 0605} 026 (2006).
%\cite{NewMuonNMC:1990xyw}
\bibitem{nmc}
D.~Allasia {\it et al.} (NMC),
%``Measurement of the neutron and the proton F2 structure function ratio,''
\emph{Phys. Lett. B} {\bf249}, 366-372 (1990).
%\cite{SeaQuest:2021zxb}
\bibitem{seaquest}
J.~Dove {\it et al.} (SeaQuest),
%``The asymmetry of antimatter in the proton,''
\emph{Nature} {\bf 590}, 561-565 (2021)
[erratum: \emph{Nature} {\bf 604}, E26 (2022)].
%\cite{STAR:2020vuq}
\bibitem{starwr}
J.~Adam \textit{et al.} (STAR),
%``Measurements of $W$ and $Z/\gamma^*$ cross sections and their ratios in p+p collisions at RHIC,''
\emph{Phys. Rev. D} {\bf 103}, 012001 (2021).
%[arXiv:2011.04708 [nucl-ex]].
%19 citations counted in INSPIRE as of 19 Nov 2024
%\cite{Nam:2021wpi}
\bibitem{wrprelim}
J.~D.~Nam for the STAR Collaboration,
%``Measurements of $W$ and $Z/γ^*$ cross sections and their ratios in $pp$ collisions at STAR,''
\emph{SciPost Phys. Proc.} {\bf 8}, 140 (2022).
%[arXiv:2108.01029 [hep-ex]].
%\cite{STAR:2021fgw}
%\cite{Hou:2019efy}
\bibitem{ct18}
T.~J.~Hou, {\it et al.}
%``New CTEQ global analysis of quantum chromodynamics with high-precision data from the LHC,''
\emph{Phys. Rev. D} {\bf 103}, 014013 (2021).
%doi:10.1103/PhysRevD.103.014013
%[arXiv:1912.10053 [hep-ph]].
%671 citations counted in INSPIRE as of 19 Nov 2024
%\cite{Bailey:2020ooq}
\bibitem{msht20}
S.~Bailey, {\it et al.}
%S.~Bailey, T.~Cridge, L.~A.~Harland-Lang, A.~D.~Martin and R.~S.~Thorne,
%``Parton distributions from LHC, HERA, Tevatron and fixed target data: MSHT20 PDFs,''
\emph{Eur. Phys. J. C} {\bf81}, 341 (2021).
%doi:10.1140/epjc/s10052-021-09057-0
%[arXiv:2012.04684 [hep-ph]].
%416 citations counted in INSPIRE as of 19 Nov 2024
\bibitem{stardipi0}
M.~S.~Abdallah {\it et al.} (STAR),
%``Evidence for Nonlinear Gluon Effects in QCD and Their Mass Number Dependence at STAR,''
\emph{Phys. Rev. Lett.} {\bf 129}, 092501 (2022).
%doi:10.1103/PhysRevLett.129.092501
%[arXiv:2111.10396 [nucl-ex]].
%24 citations counted in INSPIRE as of 19 Nov 2024
%\cite{RHICSPIN:2023zxx}
\bibitem{rhicqcd2023}
E.~C.~Aschenauer {\it et al.} (RHIC SPIN),
%``The RHIC Cold QCD Program,''
arXiv:2302.00605 [nucl-ex].
%10 citations counted in INSPIRE as of 19 Nov 2024
\end{thebibliography}
\end{document}